\documentclass[amssymb,useAMS,prd,aps,amsmath,superscriptaddress,nofootinbib,twocolumn]{revtex4}
\pdfoutput=1
\usepackage{color}
\usepackage{pslatex}
\usepackage{pdfpages}
\usepackage{graphicx}
\usepackage{psfrag}
\usepackage{pdftricks}
\usepackage{multirow}
\usepackage{graphicx}

\begin{document}
\def\mlsp{m_{\chi_1^0}}
\def\msl{M_{\tilde l}}
\def\msq{M_{\tilde q}}
\def\bsg{B(b\rightarrow s\gamma)}
\def\bsmu{B(B_s\rightarrow\mu^+\mu^-)}
\def\btau{R(B\rightarrow\tau\nu)}
\def\omg{\Omega h^2}
\def\sip{\sigma^{SI}_{\chi p}}
\def\amu{\delta a_\mu}
\def\gmuon{(g-2)_\mu}
\def\lsp{\tilde\chi^0_1}
\def\mse{m_{\tilde{e}}}
\def\msmu{m_{\tilde{\mu}}}
\def\mstau{m_{\tilde{l}_1}}
\setlength{\parindent}{0pt}

\title{Revisiting light neutralino scenarios in the MSSM}

\author{Daniel Albornoz V\'asquez}
\affiliation{LAPTH, U. de Savoie, CNRS, BP 110,
 74941 Annecy-Le-Vieux, France.}

\author{Genevi\`eve B\'elanger}
\affiliation{LAPTH, U. de Savoie, CNRS, BP 110,
 74941 Annecy-Le-Vieux, France.}

\author{C\'eline B\oe hm}
\affiliation{LAPTH, U. de Savoie, CNRS, BP 110,
 74941 Annecy-Le-Vieux, France.}
\affiliation{IPPP, Ogden centre, Durham University, UK}

\date{today}

\begin{abstract}
We revisit the case of a light neutralino LSP in the framework of the MSSM. We consider a model with eleven free parameters. We show that all scenarios where the annihilation of light neutralinos rely mainly on the exchange of a light pseudoscalar are excluded by direct detection searches and by Fermi measurements of the $\gamma-$flux from dwarf spheroidal galaxies. On the other hand, we find scenarios with light sleptons that satisfy all collider and astroparticle physics constraints. In this case, the lower limit on the LSP mass is 12.6 GeV. We discuss how the parameter space of the model will be further probed by new physics searches at the LHC.
\end{abstract}

\maketitle

\section{Introduction}

Motivated by the possible annual modulation signals reported by 
DAMA~\cite{Bernabei:2010mq} and CoGeNT~\cite{Aalseth:2010vx} that are compatible with a light dark matter candidate, several groups have recently re-investigated supersymmetric scenarios with a LSP around 10-30 GeV. Within the context of the MSSM, such a light neutralino LSP can only be obtained by relaxing the unification condition on the gaugino masses such that $M_1 \ll M_2$~\cite{Dreiner:2009ic, Fornengo:2010mk,Scopel:2011qt, Calibbi:2011ug, Belikov:2010yi, Cumberbatch:2011jp, Cao:2010fi, Vasquez:2010ru}. When this condition is satisfied, most of the stringent LEP constraints (in particular the limit of 46 GeV~\cite{Nakamura:2010zzi} on the lightest neutralino obtained within the framework of the CMSSM) are removed. Only the LEP constraints from the invisible width of the Z boson and the production of the LSP with one of the heavier neutralinos remain in the neutralino sector. Light neutralinos DM are then constrained mainly by the relic density measurement as well as by the Higgs and the flavour sector, in particular from B-physics. 

For very light neutralinos, with a mass below 10 GeV, the WMAP upper bound can be satisfied when $M_A$, the pseudoscalar mass, is around 100-150 GeV and $\tan\beta$ is large. This corresponds to a set of parameters that leads to large deviations in the flavour sector. Furthermore, this range of parameters faces severe constraints from collider searches for the heavy Higgs doublet produced in association with b quarks and decaying into tau pairs at the Tevatron and the LHC~\cite{Benjamin:2010xb,Chatrchyan:2011nx,CMS_higgs}. Indeed, at large values of $\tan\beta$ and for a light pseudoscalar, the cross section for this process is strongly enhanced. The light neutralino scenarios that survive all collider and cosmological constraints are then challenged by the upper limits from direct detection searches~\cite{Vasquez:2010ru}. Light neutralinos can also lead to strong signals in indirect searches both for the photon flux~\cite{Vasquez:2011js} or for the antiproton flux~\cite{Lavalle:2010yw, Cerdeno:2011}.

For heavier neutralinos the WMAP condition can also be satisfied when pair of LSP's annihilate into fermions through sfermion exchange or through Z exchange. The former requires light sleptons -these are however subject to LEP constraints- while the latter is efficient when the mass of the LSP pair approaches the Z mass. In either case, a light pseudoscalar is no longer needed and such scenarios can escape all constraints~\cite{Vasquez:2010ru,Dreiner:2009ic, Cumberbatch:2011jp,Gogoladze:2009mc}.

In Ref.~\cite{Vasquez:2010ru}, an exploration of the MSSM with eight free parameters based on an MCMC approach, showed that neutralino candidates below 15 GeV were severely constrained by Higgs searches as well as by direct detection searches, in particular by XENON100~\cite{Aprile:2010um}. Such a conclusion, was challenged in~\cite{Fornengo:2010mk,Calibbi:2011ug} as well as very recently in~\cite{Cumberbatch:2011jp}. 

In view of these results, we revisit in this paper the case of light neutralinos in the MSSM. We have extended our previous study in several ways. First, we increase the number of free parameters of the model to eleven, adding the gluino mass, decorrelating the third generation squark masses from the other two and splitting the left- and right-handed slepton masses. The first two parameters affect mainly the flavour sector and do not impact directly on the very light neutralino candidates. Splitting the slepton mass and exploring carefully the region where sleptons are just above the LEP limit allows to find new scenarios where neutralinos annihilate via light sleptons exchange. Second, we have also updated the limits on the Higgs sector, in particular exploring more carefully the region where all Higgses are around 100 GeV. This allows to find new scenarios with $\mlsp \lesssim 15$~GeV satisfying the Tevatron limits on the search for the heavy Higgs doublet. However we show that recent CMS results~\cite{Chatrchyan:2011nx,CMS_higgs} for Higgs searches at the LHC further constrain a large fraction of the light neutralino scenarios and that all these scenarios are excluded by the XENON100 experiment. We have also computed the fluxes of gamma rays from dwarf spheroidal galaxies (dSph) and used the limits from Fermi-LAT to constrain all scenarios with a light pseudoscalar. We find a lower bound on the neutralino mass 12.6 GeV after including the updated constraints from both colliders and astroparticle physics. These light neutralinos are always associated with sleptons just above the LEP limit. We also briefly discuss how these results will be affected by upcoming LHC results on sparticle searches and on B-physics observables.

This paper is organized as follows: in Section II, we summarize the various constraints on the model, in section III we find the lower bound on the neutralino mass and emphasize the impact of astrophysical constraints. In section IV we discuss implications for LHC observables. A discussion and comparison with other studies is presented in section V.

\section{Constraints on light neutralinos}

We consider the MSSM with eleven free parameters defined at the electroweak scale, as listed in Table~\ref{tab:MSSM_M15_parameters_2} and we assume $A_b=A_\tau=0$. These parameters only play a role in the mixing in the down sector ($\propto A_{b(\tau)}-\mu\tan\beta$), while a large mixing can be induced by $\mu\tan\beta$. To explore the parameter space we have used the Markov Chain Monte-Carlo code presented in \cite{Vasquez:2010ru} which is based on micrOMEGAs~\cite{Belanger:2005kh,Belanger:2008sj,Belanger:2010gh} for the computation of collider and flavour constraints as well as for dark matter observables. We rely on Suspect~\cite{Djouadi:2002ze} for the computation of the spectrum. The constraints imposed are listed in Table I of Ref.~\cite{Vasquez:2010ru}. They include the WMAP constraint on the abundance of dark matter~\cite{Komatsu:2008hk}, branching ratios for $\bsg, \bsmu,\btau$, the muon anomalous magnetic moment, $\gmuon$ as well as LEP limits on sparticle masses, on the invisible width of the Z and on the associated production of the LSP with a heavier neutralino. For the LEP limits we have used the values implemented in micrOMEGAs, corresponding in particular to the values for the sleptons, $\mse >100~{\rm GeV}$, $\msmu> 99~{\rm GeV}$, $\mstau >80.5~{\rm GeV}$ and $m_{\tilde\nu}>43~{\rm GeV}$.~\footnote{We have not included the flavour constraint from $K\rightarrow l\nu$, although constraining the light charged Higgs as shown in~\cite{Calibbi:2011ug}, this has no direct influence on the light neutralino.}

In this analysis we have replaced the limit on the light Higgs mass with improved limits on the Higgs sector obtained from the HiggsBounds3.1.3 package~\cite{Bechtle:2008jh,Bechtle:2011sb} linked to micrOMEGAs2.4. In this way, we take into account both the LEP constraints on the light Higgs as well as Tevatron constraints on heavy Higgs searches at large $\tan\beta$. The likelihood for the Higgs constraint is taken to be 0 when a point is rejected by HiggsBounds and 1 otherwise. We compute the global weight $\mathcal{Q}$ by multiplying the global likelihood to the global prior of each scenario. We use the likelihood and prior functions described in~\cite{Vasquez:2010ru}.

\begin{table}[hbt]
\centering
\begin{tabular}{|c|c|c|c|}
\hline
&&&\\
\rm{Parameter} & \rm{Minimum} & \rm{Maximum} & \rm{Tolerance} \\
&&&\\
\hline
$M_1$ & 1 & 1000 & 3 \\
$M_2$ & 100 & 2000 & 30 \\
$M_3$ & 500 & 6500 & 10 \\
$\mu$ & 0.5 & 1000 & 0.1 \\
$\tan\beta$ & 1 & 75 & 0.01 \\
$M_A$ & 1 & 2000 & 4 \\
$A_t$ & -3000 & 3000 & 100 \\
$M_{\tilde{l}_R}$ & 70& 2000 & 15 \\
$M_{\tilde{l}_L}$ & 70 & 2000 & 15 \\
$M_{\tilde{q}_{1,\,2}}$ & 300 & 2000 & 14 \\
$M_{\tilde{q}_3}$ & 300 & 2000 & 14 \\
\hline
\end{tabular}
\caption{Intervals for MSSM free parameters (GeV units).}
\label{tab:MSSM_M15_parameters_2}
\end{table}

We have not included recent LHC results on heavy Higgs searches~\cite{Chatrchyan:2011nx,CMS_higgs} in the fit but impose them a posteriori. Also we have not included the recent results from the LHC on squarks and gluino searches as they are somewhat model dependent. Note that when imposing cosmological constraints we allow for the possibility that neutralinos do not explain all of the dark matter in the universe but only a fraction taken to be as small as 10\%, this has no major impact on our conclusions since light neutralinos tend to be over abundant.

Light neutralinos can also be constrained by direct and indirect detection. We will apply these constraints only after having selected the best scenarios from a global fit. Specifically we will consider the XENON100 results from direct detection searches. In all cases with two scalar Higgses with a mass around 100 GeV that must couple sufficiently to the LSP to provide enough annihilation in the early universe, we expect an important contribution of both Higgses to the spin independent neutralino nucleon elastic scatttering cross section. This will turn out to be an important constraint on light neutralinos as will be discussed in the next section. 

Pair annihilation of neutralino DM into quarks and/or $\tau$'s leads, after hadronization, to the production of gamma-rays. Photons can also be radiated directly from an internal line or from a final state before it decays. The photon flux is proportional to $1/\mlsp^2$ thus a large flux is expected for light dark matter. The observation of the photon flux from dwarf spheroidal galaxies (dSph) by Fermi-LAT therefore provides a constraint on light neutralino dark matter. For each viable scenario found by the MCMC, we have computed the gamma ray flux expected in the eight dwarfs observed by the Fermi experiment. This value is then compared with the Fermi-LAT 95\% limits~\cite{Abdo:2010ex} with the procedure described in~\cite{Vasquez:2011js}. The most stringent limits are obtained for the Draco dSph.

\section{The lower limit on the neutralino mass}

Viable scenarios with light neutralinos can be difficult to find. Therefore, we have imposed the prior $\mlsp<30$~GeV. Since we already know that there are neutralinos at around $\sim 28\, GeV$~\cite{Vasquez:2010ru}, there is no need to probe higher masses, which would make the run less efficient.

\begin{figure}[htb]
\centering		\includegraphics[scale=0.12,natwidth=3cm,natheight=3cm]{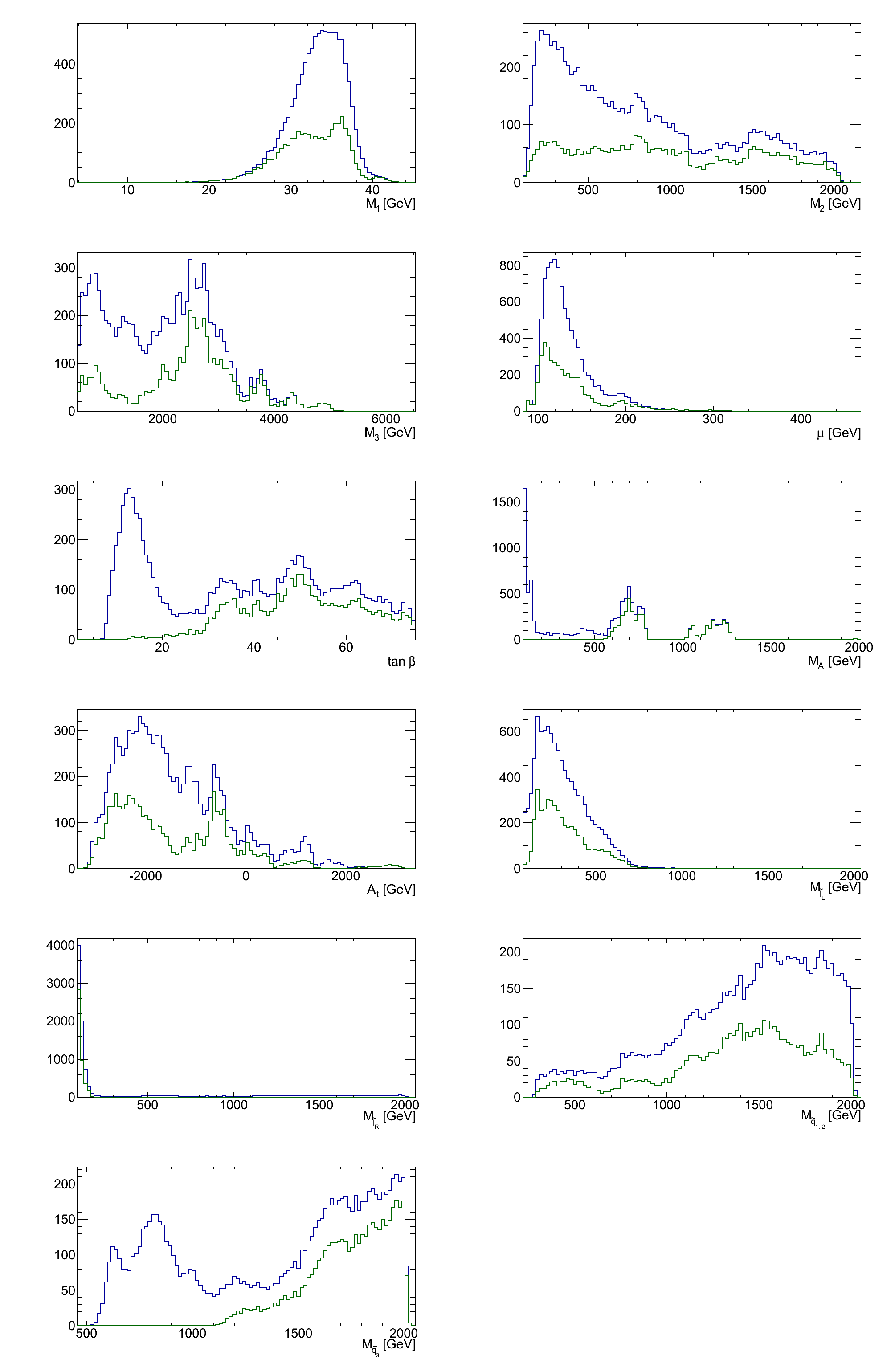}
\caption{Frequency distributions of free parameters in the light MSSM neutralino scenarios. Blue curves contain all allowed points while green curves show the distribution for the points that pass all astroparticle physics constraints. }
\label{fig:MSSM_M30_ad_parameters}
\end{figure}

Performing the MCMC analysis, we found the maximum weight to be ${\cal Q}_{\rm max}\simeq0.72$. Nevertheless, only 2.9\% of the points have weights $\mathcal{Q} \geq 0.23$ ($1\sigma$ away from ${\cal Q}_{\rm max}$), while 57\% have weights $\mathcal{Q} \geq 2.2\times10^{-3}$ ($3\sigma$ away from ${\cal Q}_{\rm max}$). We find neutralinos with masses as low as 10.5~GeV, although most points are located near 30~GeV, the prior upper bound on the neutralino mass. The allowed parameter space, represented in Fig.~\ref{fig:MSSM_M30_ad_parameters}, is best described in terms of the properties of the neutralinos that satisfy the relic density upper limit. There are three dominant mechanisms that provide efficient neutralino annihilation: A) annihilation into lepton pairs through slepton exchange, B) annihilation via exchange of a light pseudoscalar Higgs, C) annihilation via a $Z$ boson. The latter works better for masses near $M_Z/2$, therefore neutralinos below $\simeq 25$~GeV are expected to correspond to scenarios A and B.

The first scenarios (A) require a bino LSP and light sleptons, in particular right-handed sleptons which couple more strongly to the bino. Consequently, we observe a large peak at low values of the soft parameter, $m_{{\tilde l}_R}$. Furthermore, large values of $\tan\beta$ will induce a large mixing in the stau sector, thus decreasing the lightest stau mass. As a result, all sleptons are just above the LEP exclusion region. The $\tan\beta$ distribution thus extends to the highest values probed. The other two scenarios (B and C) require a LSP with as large as possible higgsino component to ensure sufficient coupling to the $Z$ or the Higgs -this means small $\mu$- even though the LSP is dominantly bino since $M_1\ll \mu$. Scenario B further requires a light pseudoscalar, hence the large peak in the distribution at low values of $M_A$. In this case, large values of $\tan\beta$ also are needed for efficient annihilation. However the low $M_A$ - large $\tan\beta$ region is strongly constrained by Tevatron searches. Furthermore the $R(B_u \rightarrow \tau \nu_{\tau})$ ratio in the case of a light charged Higgs drops to very low values around $\tan\beta=25$, thus these values are disfavored. Other parameters are constrained from several observables. For example, a squark contribution is needed to cancel the Higgs contribution in the $\bsg$, hence the peak at low values of third generation squark masses $M_{\tilde{q}_3}$. This is relevant only for scenario B.

\begin{figure}[htb]
\centering		\includegraphics[scale=0.24,natwidth=6cm,natheight=6cm]{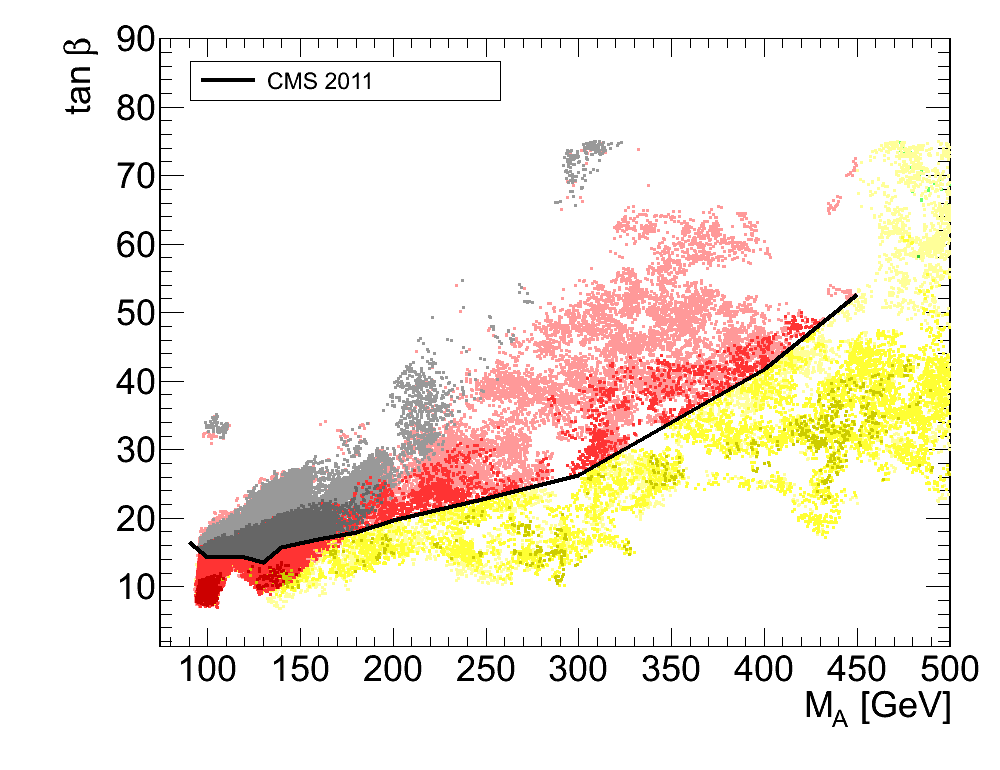}
\caption{Allowed points in the $\tan\beta$ vs. $M_A$ plane in the $\mlsp < 30$~GeV search. We show only the region where $M_A<500$~GeV. The exclusion limit from CMS is also displayed. In yellow (red), points excluded by one (two) constraint and in black those excluded by three constraints (CMS, XENON100 and dSph as described in section~\ref{sec:astro}). The shading represents $\mathcal{Q}$: weights of darker points are at most at $1\sigma$ from $\mathcal{Q}_{\rm max}$ while the lighter points are at most at $2\sigma$ and $3\sigma$.}
\label{fig:MSSM_M30_ad_tanb_vs_mA}
\end{figure}

The allowed region displayed in the $\tan\beta-M_A$ plane, Fig.~\ref{fig:MSSM_M30_ad_tanb_vs_mA}, shows that when the pseudoscalar is light, large values of $\tan\beta$ are ruled out after taking into account Tevatron constraints on Higgs decaying into tau pairs. Furthermore, the newer exclusion limit from CMS~\cite{CMS_higgs} in the same channel (black line in Fig.~\ref{fig:MSSM_M30_ad_tanb_vs_mA}) further cuts into the parameter space, the only remaining points for $M_A<150$~GeV correspond to $\tan\beta \leq 14$. 

\begin{figure}[htb]
\centering		\includegraphics[scale=0.24,natwidth=6cm,natheight=6cm]{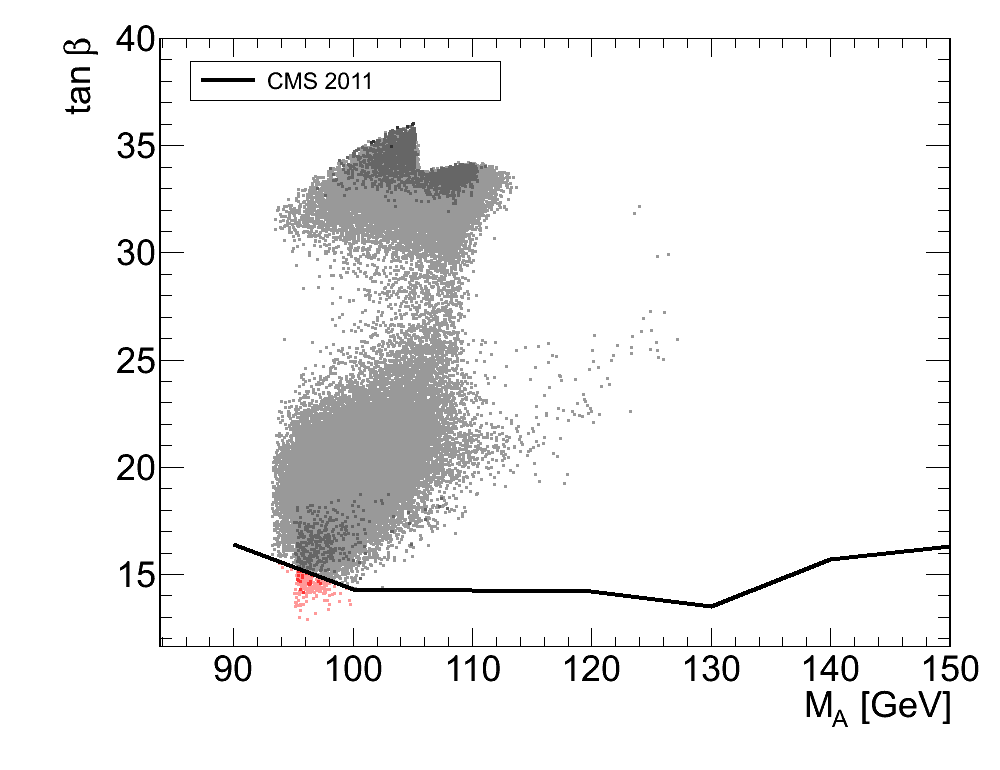}
\caption{Allowed points in the $\tan\beta$ vs. $M_A$ plane with the prior $\mlsp<15$~GeV showing only the region where $M_A<150$~GeV. The exclusion limit from CMS is also displayed. The color code is the same as in Fig.~\ref{fig:MSSM_M30_ad_tanb_vs_mA}}
\label{fig:MSSM_M15_ad_tanb_vs_mA}
\end{figure}

To ensure that we have probed completely light neutralino scenarios, we did a further run imposing a prior $\mlsp<15$~GeV. The maximum weight in that region is of $0.22$, and the maximum weight for the points with $M_A<150$~GeV is ${\cal Q}=0.085$ which is much lower than in the previous sample. The allowed points in the $M_A-\tan\beta$ are displayed in Fig.~\ref{fig:MSSM_M15_ad_tanb_vs_mA}, here we show only the region with a light pseudoscalar, other points with very large values of $\tan\beta$ and light sleptons were also found and will be discussed below. With the incorporation of the latest Tevatron bounds, we have not found the same configurations as in our previous analysis~\cite{Vasquez:2010ru}. Those with large values of $\tan\beta$ are now ruled out by Higgs searches. We found more scenarios where all Higgs bosons are around 100 GeV, indeed all Higgs bosons have to be light in order to overcome the limits on the Higgs mass from LEP. However most of these points are now constrained by the latest CMS exclusion~\cite{CMS_higgs} imposed after performing the fit. In this sample, there were neutralinos below 10 GeV that passed all collider constraints, albeit with a low weight. However, we will show that all these neutralinos are ruled out by astroparticle limits (both from XENON100 and Fermi-LAT).

\subsection{Constraints from astroparticle physics}
\label{sec:astro}

\begin{figure}[htb]
\centering		\includegraphics[scale=0.24,natwidth=6cm,natheight=6cm]{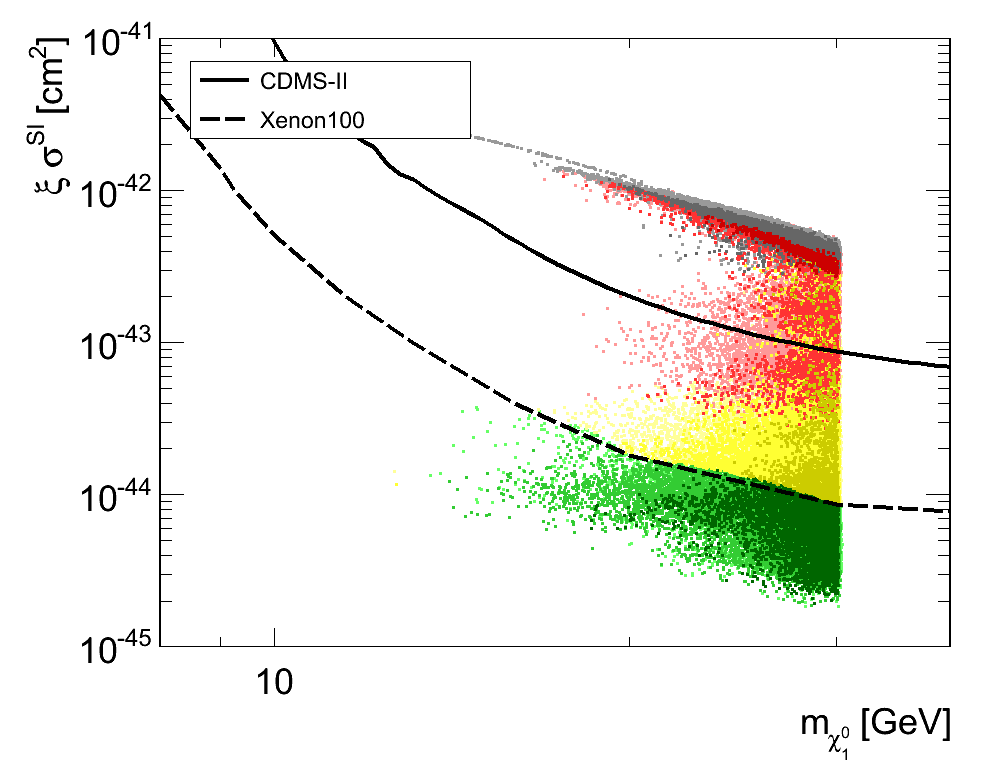}
\caption{Points of the $\mlsp < 30$~GeV search represented in the $\xi \sigma^{SI}$ vs. neutralino mass plane. Exclusion limits from CDMS-II~\cite{Ahmed:2009zw} and XENON100 are shown. The color code is the same as in Fig.~\ref{fig:MSSM_M30_ad_tanb_vs_mA}, green points are allowed.}
\label{fig:MSSM_M30_ad_SI}
\end{figure}

We now consider two different astroparticle constraints on the light neutralino scenarios. First we consider the spin independent direct detection limits from XENON100 as it provides the most stringent limit on light neutralinos. Figure~\ref{fig:MSSM_M30_ad_SI} represents the yields in the $\xi \sigma^{SI}$ vs. $\mlsp$ plane along with limits from XENON100 and CDMS-II. The three types of scenarios have very different predictions for the spin independent cross section on nucleons. In scenario A, the LSP can be pure bino and therefore couples weakly to the Higgs, cross sections can therefore be much suppressed. It is in this class of scenarios (green points) that we find the lightest viable neutralino. In case B, cross sections which receive a contribution from both light scalar Higgses are large. All these points are ruled out by XENON100 as was found in the previous analysis. In scenario C, the LSP also has a higgsino component but tends to have a lower cross section on nucleons since it receives only the contribution of one light Higgs. Since a smaller higgsino component is needed as one approaches the $Z$ resonance, some of these scenarios with mass near 30 GeV predict an elastic scattering cross section below the limit of XENON100. When computing these predictions, we have chosen rather conservative values for the quark coefficient in the nucleon ($\sigma_{\pi N}=45$~MeV, $\sigma_0=40$~MeV) although recent lattice QCD results~\cite{Giedt:2009mr} indicate that the s-quark content could be smaller than previously thought, leading to a suppression of the SI cross sections. Taking the central value from the result would only lead to a 20\% further reduction in the neutralino proton cross section. This is not enough to make some of the scenarios B drop below the XENON100 exclusion limit. An improvement on the SI direct detection limit by a factor 4-8 is required to close light neutralino scenarios up to 30 GeV.

We now compute the gamma ray flux originating from DM annihilation in dSph assuming an NFW~\cite{Navarro:1996gj} profile and compare this with the 95\% limits from Fermi-LAT considering an angular region of $0.5^\circ$ and an integrated flux from $0.1{\rm GeV} < E< m_{\tilde{\chi_1^0}}$. We found that many configurations -more specifically with the characteristics of scenario B- are excluded by both limits (red points) while others are also constrained by XENON100 (yellow points). The configurations allowed by XENON100 (green points) satisfy all collider and astrophysical constraints. All these belong to scenario A for neutralinos below 28 GeV. In these configurations, the photon flux just reached the maximal value allowed by Fermi for the smallest mass (recall that the flux goes as $1/\mlsp^2$). 

\begin{figure}[htb]
\centering		\includegraphics[scale=0.24,natwidth=6cm,natheight=6cm]{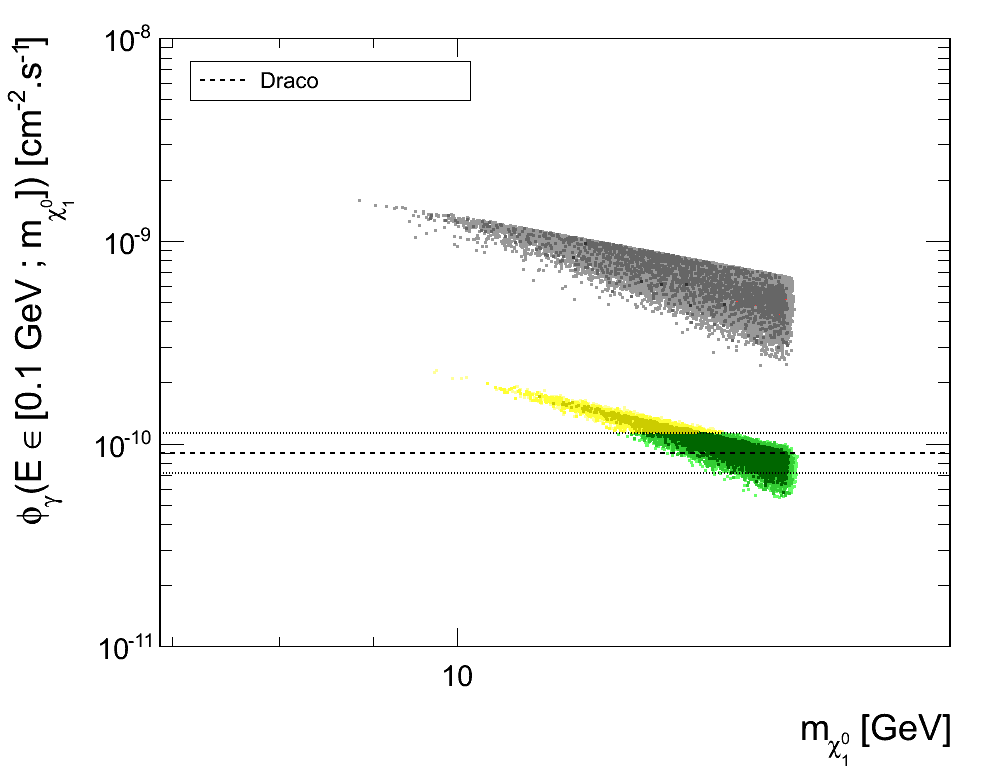}
\caption{Integrated $\gamma$-ray flux from the Draco dwarf spheroidal galaxy as a function of the neutralino mass in the $\mlsp<15$~GeV search. We show limits from Fermi-LAT. Same color code as Fig.~\ref{fig:MSSM_M30_ad_SI}.}
\label{fig:MSSM_M15_ad_Dwarf_vs_Mchi_Draco}
\end{figure}

Since the parameter space allowing for light neutralinos is rather fine-tuned, one might argue that somewhat lighter neutralinos could be found with a refined analysis. With an additional run with a prior set at $m_{\tilde\chi}<15$~GeV, we found that the lower bound on the neutralino could be extended by a few GeV's when considering collider constraints. However the lighter neutralinos were constrained by dSphs as displayed in Fig.~\ref{fig:MSSM_M15_ad_Dwarf_vs_Mchi_Draco}. In this run we found the lower limit on the neutralino mass to be 12.6 GeV, corresponding to a point of weight $\mathcal{Q}\simeq 0.11$ that is safe regarding XENON100, Fermi-LAT and CMS. As before, it corresponds to scenarios with light sleptons.

\begin{figure}[htb]
\centering		\includegraphics[scale=0.24,natwidth=6cm,natheight=6cm]{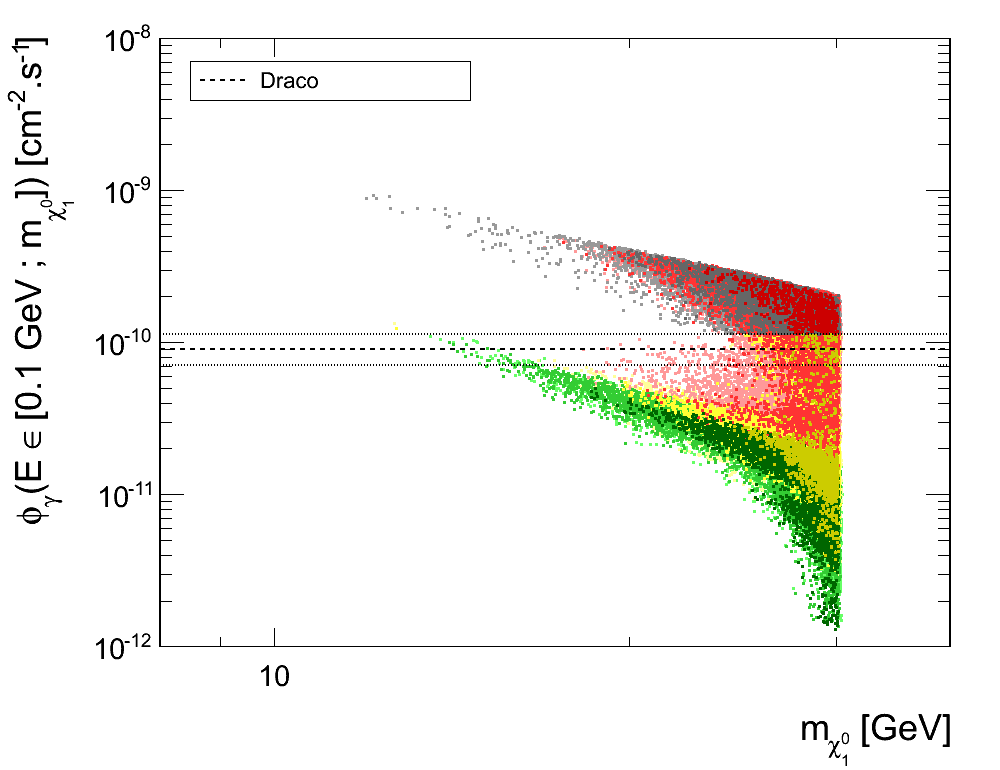}
\caption{Integrated $\gamma$-ray flux from the Draco dwarf spheroidal galaxy as a function of the neutralino mass in the $\mlsp < 30$~GeV search. We show limits from Fermi-LAT. Same color code as Fig.~\ref{fig:MSSM_M30_ad_SI}.}
\label{fig:MSSM_M30_ad_Dwarf_vs_Mchi_Draco}
\end{figure}

After taking into account constraints from direct and indirect DM searches and considering only the points with the highest likelihood we find that the lightest neutralino has a mass of $m_{{\chi_1}^0}\simeq 18.6$~GeV, while 12.6 GeV is possible with the prior $\mlsp < 15$~GeV. Other constraints are not a critical issue as the light slepton is favorable for the muon anomalous magnetic moment and the large value for $M_A$ implies that the B-physics constraints are weak. Furthermore the almost pure bino LSP easily evades the LEP constraints on the $Z$ invisible width. These new configurations were not found in our previous study where we had assumed one common soft slepton mass, furthermore they rely critically on the exact value taken for the limit on light sleptons. These results are in qualitative agreement with the recent results of~\cite{Cumberbatch:2011jp}.

\section{Other collider observables}

\begin{figure}[htb]
\centering		\includegraphics[scale=0.24,natwidth=6cm,natheight=6cm]{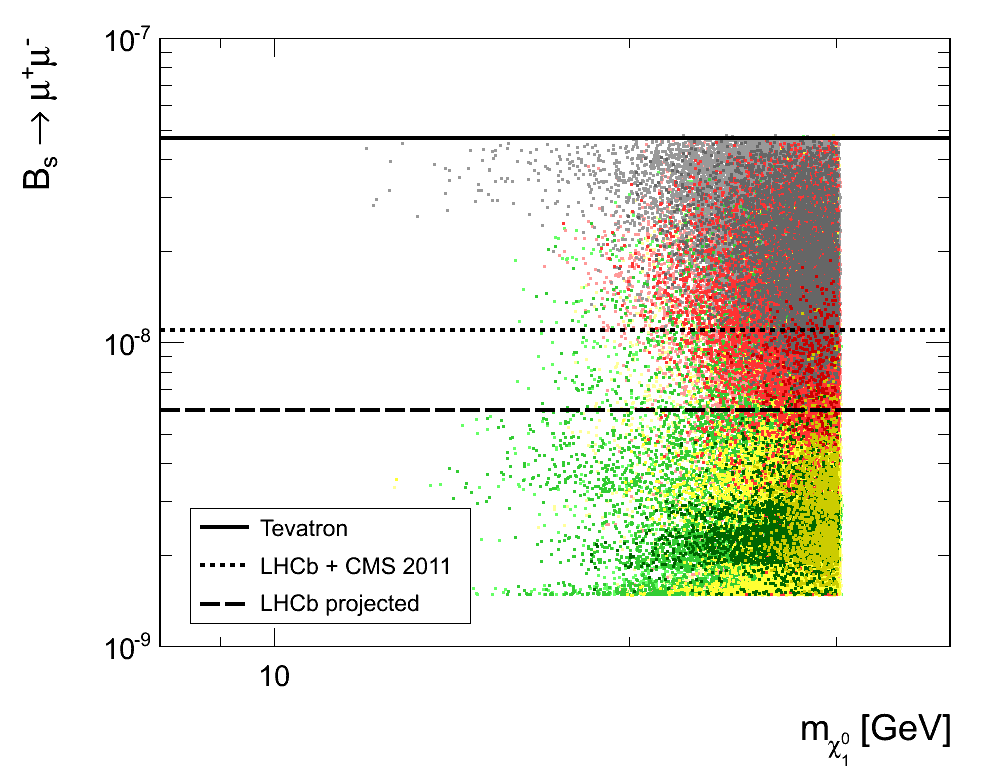}
\caption{Predictions for $Br(B_s\rightarrow \mu^+\mu^-)$ as a function of the LSP mass in the $\mlsp<30$~GeV search. The current Tevatron limit (full), the combined LHCb and CMS limit (dot)~\cite{bsmu_CMSLHCB} as well as the projected LHCb limit (dash)~\cite{lhcb} are also displayed. The color code is the same as in Fig.~\ref{fig:MSSM_M30_ad_SI}}.
\label{fig:MSSM_M30_ad_bsmumu_vs_mchi}
\end{figure}

We now consider the prospects for probing these scenarios at the LHC. For this, we have computed the value for all the observables used for the fit as well as the masses of sparticles. One observable that is promising in the flavour sector is $\bsmu$, since it is enhanced at large values of $\tan\beta$ and low values of $M_A$. Even though this region is already constrained from Higgs searches, the predictions for $\bsmu$ together with the recent limit obtained from a combination of LHCb and CMS results~\cite{bsmu_CMSLHCB} as well as expectations for the reach of LHCb~\cite{lhcb} show that many scenarios would be either further constrained or lead to a signal in the very near future, see Fig.~\ref{fig:MSSM_M30_ad_bsmumu_vs_mchi}. However most of the configurations with the best likelihood with neutralinos below 20 GeV predict a rate much below the foreseen limit. These all belong to the scenarios with light sleptons.

As mentioned above, we have not imposed the LHC constraints on squarks and gluinos in the MCMC analysis. We have however checked a posteriori that these constraints did not impact the lower limit on the neutralino mass. For this we have used the limits set by ATLAS, $m_{\tilde q}>850$~GeV and $ m_{\tilde g}> 800$~GeV~\cite{ATLAS-susy}, in a simplified model where the squarks of the first generations are degenerate and assumed to decay uniquely in jets plus missing energy. In our case the limits are somewhat weaker as the squarks have reduced branching ratios in jets plus missing energy. In any case, many of the scenarios with a good likelihood have first generation squarks and/or gluinos above the TeV scale (as indicated by the soft mass distributions in Fig.~\ref{fig:MSSM_M30_ad_parameters}). In particular many scenarios with the best likelihoods have $m_{\tilde g}>2$~TeV, that is above the mass range that can be probed with the high energy, high luminosity LHC ($\sqrt{s}=14$~TeV and ${\cal L}=100 {\rm fb}^{-1}$). This is not surprising since the color sector affects only the light neutralino scenarios through some of the B-physics observables.

\begin{figure}[htb]
\centering		\includegraphics[scale=0.8,natwidth=6cm,natheight=6cm]{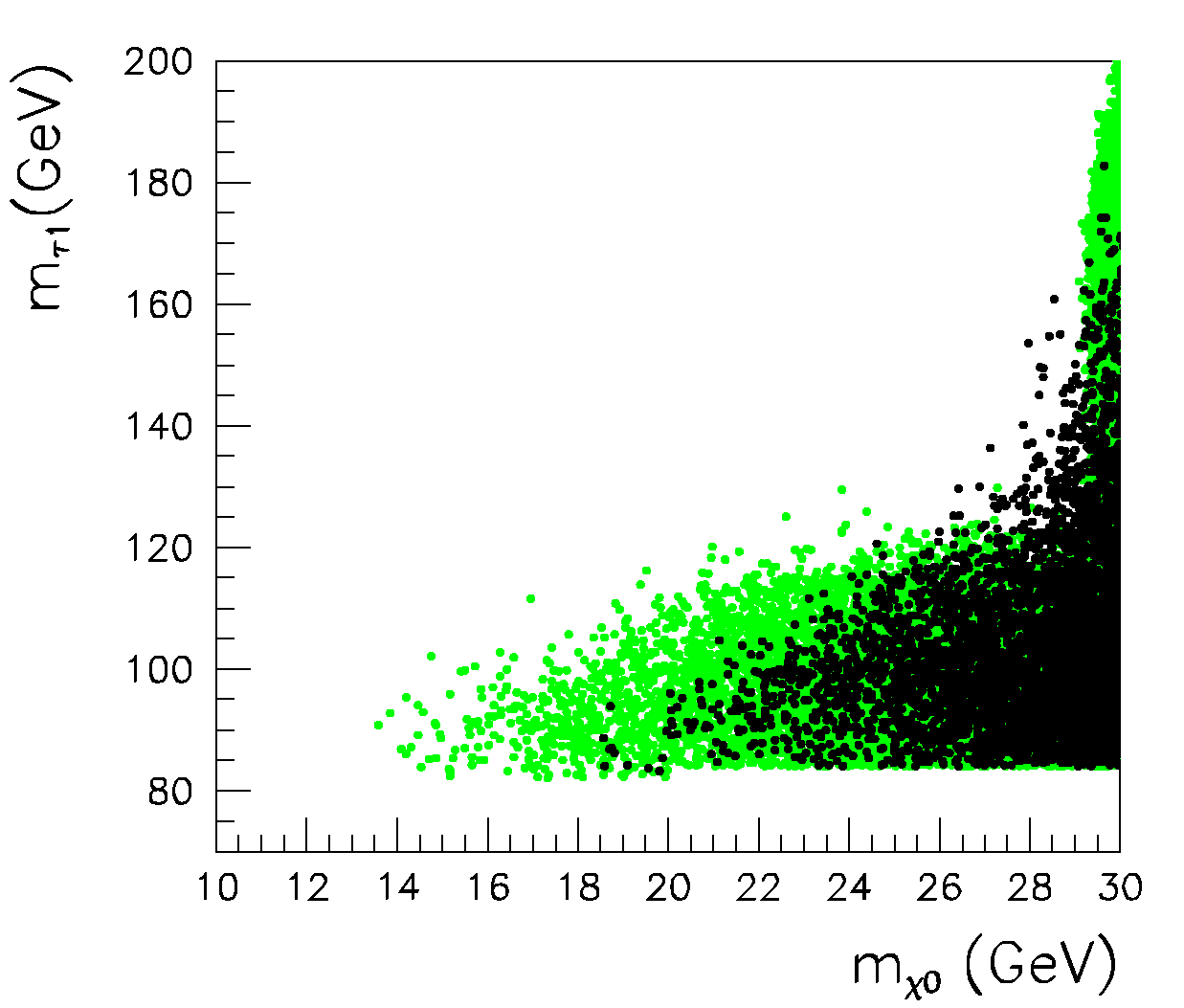}
\caption{Predictions for the lightest slepton mass as a function of the LSP mass for points that pass all collider and astroparticle physics constraints in the $\mlsp<30$~GeV search. Dark points are those within the $1\sigma$ region, and light points within $3\sigma$.}
\label{fig:MSSM_M30_mslepton_vs_mneutralino}
\end{figure}

The points that survive all collider and astrophysical limits nevertheless predict some light particles and can therefore be probed further at the LHC. These points are displayed in the $m_{\tilde\tau}-m_{\tilde\chi}$ plane in Fig.~\ref{fig:MSSM_M30_mslepton_vs_mneutralino} where it is shown that sleptons with a mass below 120 GeV are predicted in all scenarios where the LSP is below 26 GeV. The slepton pair production cross section at the LHC-7TeV is around 20-50 fb and leads to a signature with two leptons and missing energy. These scenarios can easily be studied at a future linear collider~\cite{Conley:2010jk}. Finally the points with a neutralino near 30 GeV that belong to scenario C (annihilation through a $Z$ exchange) and that also survive all constraints predict the chargino and the second neutralino to be below 200 GeV. Both particles often decay into a LSP and a gauge boson rather than into leptons, hence one cannot exploit the limits from the trilepton searches set by the Tevatron~\cite{Abazov:2009zi,Forrest:2009gm}.

\section{Discussion}

This analysis shows that there are two windows of configurations with light neutralinos with very different characteristics and signatures. In the first, it is the light pseudoscalar Higgs boson that dominates DM neutralino annihilation. A large fraction of these configurations have been excluded by the recent results from CMS on searches for associated Higgs production decaying into $\tau$ pairs. Furthermore, their astroparticle signatures, specifically in the SI direct detection experiments and in the $\gamma$-ray signal from dSphs, are largely above the observations published by the XENON100 and the Fermi-LAT collaborations. In the second, light sleptons just above the LEP limit are required to provide efficient neutralino annihilations. The viable configurations feature a heavy pseudoscalar mass ($M_A\ge 500$~GeV) thus the heavy Higgs doublet does not contribute significantly to the SI elastic scattering of neutralinos off nucleons. This, combined with the fact that the LSP is a bino, allows to escape the direct detection constraint. These scenarios are safe regarding indirect detection limits. The absolute lower bound on the light neutralino is found to be 12.6~GeV.

The results presented here appear to disagree somewhat with~\cite{Calibbi:2011ug, Fornengo:2010mk} which obtain neutralinos lighter than 10 GeV accompanied with light pseudoscalars after taking into account collider limits. This could be explained by the fact that there are some differences as to the constraints taken into account -we used a slightly more stringent constraint on the Higgs sector and on $\bsmu$ and $\btau$ as compared to~\cite{Fornengo:2010mk}- and/or on the definition of the allowed points -combined likelihood in our case and 2$\sigma$ bound in other works. To verify this we have also performed a random scan of the parameter space using the procedure described in~\cite{Calibbi:2011ug} and found similar results. In any case, we claim that all these scenarios are constrained by dSphs and by direct detection. The absolute lower bound on the light neutralino that we find is in agreement with the one found in Ref.~\cite{Dreiner:2009ic} and in~\cite{Cumberbatch:2011jp}. In these analyses, light sleptons are required to ensure sufficient annihilation of the bino LSP.

In conclusion, the light neutralino scenarios that are allowed by collider constraints are in many case challenged by direct and indirect detection limits. The remaining scenarios do not have a large scattering cross section on nuclei and cannot explain DAMA~\cite{Bernabei:2010mq} and CoGeNT~\cite{Aalseth:2010vx}. An improvement of less than an order of magnitude  on the spin independent  direct detection  limit  would allow to rule out all MSSM neutralinos lighter than 30 GeV while a factor of two improvement in sensitivity in gamma ray searches would probe all neutralinos lighter than 20 GeV. Furthermore light neutralinos are necessarily accompanied by other new particles at the electroweak scale. In particular one expects sleptons with a mass ${\cal O}(100 GeV)$ when neutralinos are lighter than 26 GeV and/or charginos and other neutralinos with masses of ${\cal O}(200)$ GeV. an an order of magnitude on the spin independent direct detection limit would allow to rule out all MSSM neutralinos lighter than 30 GeV while a factor of two improvement in sensitivity in gamma ray searches would probe all neutralinos lighter than 20 GeV. We therefore expect the lower limit on the lightest neutralino mass to keep increasing as the LHC and direct and indirect dark matter searches improve their sensitivity.

\section{Acknowledgments}
We thank Sasha Pukhov for providing the HiggsBounds interface with micrOMEGAs and Karina Williams for helpful discussions on HiggsBounds. 
This work was supported in part by the CNRS-PICS grant ``Propagation of low energy positrons''.

\bibliography{mssm_ad}

\begin{thebibliography}{36}
\expandafter\ifx\csname natexlab\endcsname\relax\def\natexlab#1{#1}\fi
\expandafter\ifx\csname bibnamefont\endcsname\relax
  \def\bibnamefont#1{#1}\fi
\expandafter\ifx\csname bibfnamefont\endcsname\relax
  \def\bibfnamefont#1{#1}\fi
\expandafter\ifx\csname citenamefont\endcsname\relax
  \def\citenamefont#1{#1}\fi
\expandafter\ifx\csname url\endcsname\relax
  \def\url#1{\texttt{#1}}\fi
\expandafter\ifx\csname urlprefix\endcsname\relax\def\urlprefix{URL }\fi
\providecommand{\bibinfo}[2]{#2}
\providecommand{\eprint}[2][]{\url{#2}}

\bibitem[{\citenamefont{Bernabei et~al.}(2010)}]{Bernabei:2010mq}
\bibinfo{author}{\bibfnamefont{R.}~\bibnamefont{Bernabei}}
  \bibnamefont{et~al.}, \bibinfo{journal}{Eur. Phys. J.}
  \textbf{\bibinfo{volume}{C67}}, \bibinfo{pages}{39} (\bibinfo{year}{2010}),
  \eprint{1002.1028}.

\bibitem[{\citenamefont{Aalseth et~al.}(2010)}]{Aalseth:2010vx}
\bibinfo{author}{\bibfnamefont{C.~E.} \bibnamefont{Aalseth}}
  \bibnamefont{et~al.} (\bibinfo{collaboration}{CoGeNT})
  (\bibinfo{year}{2010}), \eprint{1002.4703}.

\bibitem[{\citenamefont{Dreiner et~al.}(2009)\citenamefont{Dreiner, Heinemeyer,
  Kittel, Langenfeld, Weber et~al.}}]{Dreiner:2009ic}
\bibinfo{author}{\bibfnamefont{H.~K.} \bibnamefont{Dreiner}},
  \bibinfo{author}{\bibfnamefont{S.}~\bibnamefont{Heinemeyer}},
  \bibinfo{author}{\bibfnamefont{O.}~\bibnamefont{Kittel}},
  \bibinfo{author}{\bibfnamefont{U.}~\bibnamefont{Langenfeld}},
  \bibinfo{author}{\bibfnamefont{A.~M.} \bibnamefont{Weber}},
  \bibnamefont{et~al.}, \bibinfo{journal}{Eur.Phys.J.}
  \textbf{\bibinfo{volume}{C62}}, \bibinfo{pages}{547} (\bibinfo{year}{2009}),
  \eprint{0901.3485}.

\bibitem[{\citenamefont{Fornengo et~al.}(2011)\citenamefont{Fornengo, Scopel,
  and Bottino}}]{Fornengo:2010mk}
\bibinfo{author}{\bibfnamefont{N.}~\bibnamefont{Fornengo}},
  \bibinfo{author}{\bibfnamefont{S.}~\bibnamefont{Scopel}}, \bibnamefont{and}
  \bibinfo{author}{\bibfnamefont{A.}~\bibnamefont{Bottino}},
  \bibinfo{journal}{Phys.Rev.} \textbf{\bibinfo{volume}{D83}},
  \bibinfo{pages}{015001} (\bibinfo{year}{2011}), \eprint{1011.4743}.

\bibitem[{\citenamefont{Scopel et~al.}(2011)\citenamefont{Scopel, Choi,
  Fornengo, and Bottino}}]{Scopel:2011qt}
\bibinfo{author}{\bibfnamefont{S.}~\bibnamefont{Scopel}},
  \bibinfo{author}{\bibfnamefont{S.}~\bibnamefont{Choi}},
  \bibinfo{author}{\bibfnamefont{N.}~\bibnamefont{Fornengo}}, \bibnamefont{and}
  \bibinfo{author}{\bibfnamefont{A.}~\bibnamefont{Bottino}},
  \bibinfo{journal}{Phys.Rev.} \textbf{\bibinfo{volume}{D83}},
  \bibinfo{pages}{095016} (\bibinfo{year}{2011}), \eprint{1102.4033}.

\bibitem[{\citenamefont{Calibbi et~al.}(2011)\citenamefont{Calibbi, Ota, and
  Takanishi}}]{Calibbi:2011ug}
\bibinfo{author}{\bibfnamefont{L.}~\bibnamefont{Calibbi}},
  \bibinfo{author}{\bibfnamefont{T.}~\bibnamefont{Ota}}, \bibnamefont{and}
  \bibinfo{author}{\bibfnamefont{Y.}~\bibnamefont{Takanishi}},
  \bibinfo{journal}{JHEP} \textbf{\bibinfo{volume}{1107}}, \bibinfo{pages}{013}
  (\bibinfo{year}{2011}), \eprint{1104.1134}.

\bibitem[{\citenamefont{Belikov et~al.}(2010)\citenamefont{Belikov, Gunion,
  Hooper, and Tait}}]{Belikov:2010yi}
\bibinfo{author}{\bibfnamefont{A.~V.} \bibnamefont{Belikov}},
  \bibinfo{author}{\bibfnamefont{J.~F.} \bibnamefont{Gunion}},
  \bibinfo{author}{\bibfnamefont{D.}~\bibnamefont{Hooper}}, \bibnamefont{and}
  \bibinfo{author}{\bibfnamefont{T.~M.} \bibnamefont{Tait}}
  (\bibinfo{year}{2010}), \eprint{1009.0549}.

\bibitem[{\citenamefont{Cumberbatch et~al.}(2011)\citenamefont{Cumberbatch,
  Lopez-Fogliani, Roszkowski, de~Austri, and Tsai}}]{Cumberbatch:2011jp}
\bibinfo{author}{\bibfnamefont{D.~T.} \bibnamefont{Cumberbatch}},
  \bibinfo{author}{\bibfnamefont{D.~E.} \bibnamefont{Lopez-Fogliani}},
  \bibinfo{author}{\bibfnamefont{L.}~\bibnamefont{Roszkowski}},
  \bibinfo{author}{\bibfnamefont{R.~R.} \bibnamefont{de~Austri}},
  \bibnamefont{and} \bibinfo{author}{\bibfnamefont{Y.-L.~S.}
  \bibnamefont{Tsai}} (\bibinfo{year}{2011}), \eprint{1107.1604}.

\bibitem[{\citenamefont{Cao et~al.}(2010)\citenamefont{Cao, Hikasa, Wang, Yang,
  and Yu}}]{Cao:2010fi}
\bibinfo{author}{\bibfnamefont{J.}~\bibnamefont{Cao}},
  \bibinfo{author}{\bibfnamefont{K.-i.} \bibnamefont{Hikasa}},
  \bibinfo{author}{\bibfnamefont{W.}~\bibnamefont{Wang}},
  \bibinfo{author}{\bibfnamefont{J.~M.} \bibnamefont{Yang}}, \bibnamefont{and}
  \bibinfo{author}{\bibfnamefont{L.-X.} \bibnamefont{Yu}},
  \bibinfo{journal}{JHEP} \textbf{\bibinfo{volume}{1007}}, \bibinfo{pages}{044}
  (\bibinfo{year}{2010}), \eprint{1005.0761}.

\bibitem[{\citenamefont{Vasquez et~al.}(2010)\citenamefont{Vasquez, Belanger,
  Boehm, Pukhov, and Silk}}]{Vasquez:2010ru}
\bibinfo{author}{\bibfnamefont{D.~A.} \bibnamefont{Vasquez}},
  \bibinfo{author}{\bibfnamefont{G.}~\bibnamefont{Belanger}},
  \bibinfo{author}{\bibfnamefont{C.}~\bibnamefont{Boehm}},
  \bibinfo{author}{\bibfnamefont{A.}~\bibnamefont{Pukhov}}, \bibnamefont{and}
  \bibinfo{author}{\bibfnamefont{J.}~\bibnamefont{Silk}},
  \bibinfo{journal}{Phys.Rev.} \textbf{\bibinfo{volume}{D82}},
  \bibinfo{pages}{115027} (\bibinfo{year}{2010}), \eprint{1009.4380}.

\bibitem[{\citenamefont{Nakamura et~al.}(2010)}]{Nakamura:2010zzi}
\bibinfo{author}{\bibfnamefont{K.}~\bibnamefont{Nakamura}} \bibnamefont{et~al.}
  (\bibinfo{collaboration}{Particle Data Group}), \bibinfo{journal}{J.Phys.G}
  \textbf{\bibinfo{volume}{G37}}, \bibinfo{pages}{075021}
  (\bibinfo{year}{2010}).

\bibitem[{\citenamefont{Benjamin et~al.}(2010)}]{Benjamin:2010xb}
\bibinfo{author}{\bibfnamefont{D.}~\bibnamefont{Benjamin}} \bibnamefont{et~al.}
  (\bibinfo{collaboration}{Tevatron New Phenomena and Higgs Working Group})
  (\bibinfo{year}{2010}), \eprint{1003.3363}.

\bibitem[{\citenamefont{Chatrchyan et~al.}(2011)}]{Chatrchyan:2011nx}
\bibinfo{author}{\bibfnamefont{S.}~\bibnamefont{Chatrchyan}}
  \bibnamefont{et~al.} (\bibinfo{collaboration}{CMS Collaboration})
  (\bibinfo{year}{2011}), \eprint{1104.1619}.

\bibitem[{\citenamefont{Collaboration.}(2011)}]{CMS_higgs}
\bibinfo{author}{\bibfnamefont{C.}~\bibnamefont{Collaboration.}}
  (\bibinfo{year}{2011}), \bibinfo{note}{cMS PAS HIG-11-011}.

\bibitem[{\citenamefont{Vasquez et~al.}(2011)\citenamefont{Vasquez, Belanger,
  and Boehm}}]{Vasquez:2011js}
\bibinfo{author}{\bibfnamefont{D.~A.} \bibnamefont{Vasquez}},
  \bibinfo{author}{\bibfnamefont{G.}~\bibnamefont{Belanger}}, \bibnamefont{and}
  \bibinfo{author}{\bibfnamefont{C.}~\bibnamefont{Boehm}}
  (\bibinfo{year}{2011}), \eprint{1107.1614}.

\bibitem[{\citenamefont{Lavalle}(2010)}]{Lavalle:2010yw}
\bibinfo{author}{\bibfnamefont{J.}~\bibnamefont{Lavalle}},
  \bibinfo{journal}{Phys.Rev.} \textbf{\bibinfo{volume}{D82}},
  \bibinfo{pages}{081302} (\bibinfo{year}{2010}), \eprint{1007.5253}.

\bibitem[{\citenamefont{Cerdeno et~al.}(2011)\citenamefont{Cerdeno, Delahaye,
  and Lavalle}}]{Cerdeno:2011}
\bibinfo{author}{\bibfnamefont{D.}~\bibnamefont{Cerdeno}},
  \bibinfo{author}{\bibfnamefont{T.}~\bibnamefont{Delahaye}}, \bibnamefont{and}
  \bibinfo{author}{\bibfnamefont{J.}~\bibnamefont{Lavalle}}
  (\bibinfo{year}{2011}), \eprint{1108.1128}.

\bibitem[{\citenamefont{Gogoladze et~al.}(2010)\citenamefont{Gogoladze, Khalid,
  Raza, and Shafi}}]{Gogoladze:2009mc}
\bibinfo{author}{\bibfnamefont{I.}~\bibnamefont{Gogoladze}},
  \bibinfo{author}{\bibfnamefont{R.}~\bibnamefont{Khalid}},
  \bibinfo{author}{\bibfnamefont{S.}~\bibnamefont{Raza}}, \bibnamefont{and}
  \bibinfo{author}{\bibfnamefont{Q.}~\bibnamefont{Shafi}},
  \bibinfo{journal}{Mod.Phys.Lett.} \textbf{\bibinfo{volume}{A25}},
  \bibinfo{pages}{3371} (\bibinfo{year}{2010}), \eprint{0912.5411}.

\bibitem[{\citenamefont{Aprile et~al.}(2010)}]{Aprile:2010um}
\bibinfo{author}{\bibfnamefont{E.}~\bibnamefont{Aprile}} \bibnamefont{et~al.}
  (\bibinfo{collaboration}{XENON100}) (\bibinfo{year}{2010}),
  \eprint{1005.0380}.

\bibitem[{\citenamefont{Belanger et~al.}(2005)\citenamefont{Belanger, Boudjema,
  Hugonie, Pukhov, and Semenov}}]{Belanger:2005kh}
\bibinfo{author}{\bibfnamefont{G.}~\bibnamefont{Belanger}},
  \bibinfo{author}{\bibfnamefont{F.}~\bibnamefont{Boudjema}},
  \bibinfo{author}{\bibfnamefont{C.}~\bibnamefont{Hugonie}},
  \bibinfo{author}{\bibfnamefont{A.}~\bibnamefont{Pukhov}}, \bibnamefont{and}
  \bibinfo{author}{\bibfnamefont{A.}~\bibnamefont{Semenov}},
  \bibinfo{journal}{JCAP} \textbf{\bibinfo{volume}{0509}}, \bibinfo{pages}{001}
  (\bibinfo{year}{2005}), \eprint{hep-ph/0505142}.

\bibitem[{\citenamefont{Belanger et~al.}(2009)\citenamefont{Belanger, Boudjema,
  Pukhov, and Semenov}}]{Belanger:2008sj}
\bibinfo{author}{\bibfnamefont{G.}~\bibnamefont{Belanger}},
  \bibinfo{author}{\bibfnamefont{F.}~\bibnamefont{Boudjema}},
  \bibinfo{author}{\bibfnamefont{A.}~\bibnamefont{Pukhov}}, \bibnamefont{and}
  \bibinfo{author}{\bibfnamefont{A.}~\bibnamefont{Semenov}},
  \bibinfo{journal}{Comput. Phys. Commun.} \textbf{\bibinfo{volume}{180}},
  \bibinfo{pages}{747} (\bibinfo{year}{2009}), \eprint{0803.2360}.

\bibitem[{\citenamefont{Belanger et~al.}(2011)\citenamefont{Belanger, Boudjema,
  Brun, Pukhov, Rosier-Lees et~al.}}]{Belanger:2010gh}
\bibinfo{author}{\bibfnamefont{G.}~\bibnamefont{Belanger}},
  \bibinfo{author}{\bibfnamefont{F.}~\bibnamefont{Boudjema}},
  \bibinfo{author}{\bibfnamefont{P.}~\bibnamefont{Brun}},
  \bibinfo{author}{\bibfnamefont{A.}~\bibnamefont{Pukhov}},
  \bibinfo{author}{\bibfnamefont{S.}~\bibnamefont{Rosier-Lees}},
  \bibnamefont{et~al.}, \bibinfo{journal}{Comput.Phys.Commun.}
  \textbf{\bibinfo{volume}{182}}, \bibinfo{pages}{842} (\bibinfo{year}{2011}),
  \eprint{1004.1092}.

\bibitem[{\citenamefont{Djouadi et~al.}(2007)\citenamefont{Djouadi, Kneur, and
  Moultaka}}]{Djouadi:2002ze}
\bibinfo{author}{\bibfnamefont{A.}~\bibnamefont{Djouadi}},
  \bibinfo{author}{\bibfnamefont{J.-L.} \bibnamefont{Kneur}}, \bibnamefont{and}
  \bibinfo{author}{\bibfnamefont{G.}~\bibnamefont{Moultaka}},
  \bibinfo{journal}{Comput. Phys. Commun.} \textbf{\bibinfo{volume}{176}},
  \bibinfo{pages}{426} (\bibinfo{year}{2007}), \eprint{hep-ph/0211331}.

\bibitem[{\citenamefont{Komatsu et~al.}(2009)}]{Komatsu:2008hk}
\bibinfo{author}{\bibfnamefont{E.}~\bibnamefont{Komatsu}} \bibnamefont{et~al.}
  (\bibinfo{collaboration}{WMAP Collaboration}),
  \bibinfo{journal}{Astrophys.J.Suppl.} \textbf{\bibinfo{volume}{180}},
  \bibinfo{pages}{330} (\bibinfo{year}{2009}), \eprint{0803.0547}.

\bibitem[{\citenamefont{Bechtle et~al.}(2010)\citenamefont{Bechtle, Brein,
  Heinemeyer, Weiglein, and Williams}}]{Bechtle:2008jh}
\bibinfo{author}{\bibfnamefont{P.}~\bibnamefont{Bechtle}},
  \bibinfo{author}{\bibfnamefont{O.}~\bibnamefont{Brein}},
  \bibinfo{author}{\bibfnamefont{S.}~\bibnamefont{Heinemeyer}},
  \bibinfo{author}{\bibfnamefont{G.}~\bibnamefont{Weiglein}}, \bibnamefont{and}
  \bibinfo{author}{\bibfnamefont{K.~E.} \bibnamefont{Williams}},
  \bibinfo{journal}{Comput.Phys.Commun.} \textbf{\bibinfo{volume}{181}},
  \bibinfo{pages}{138} (\bibinfo{year}{2010}), \eprint{0811.4169}.

\bibitem[{\citenamefont{Bechtle et~al.}(2011)\citenamefont{Bechtle, Brein,
  Heinemeyer, Weiglein, and Williams}}]{Bechtle:2011sb}
\bibinfo{author}{\bibfnamefont{P.}~\bibnamefont{Bechtle}},
  \bibinfo{author}{\bibfnamefont{O.}~\bibnamefont{Brein}},
  \bibinfo{author}{\bibfnamefont{S.}~\bibnamefont{Heinemeyer}},
  \bibinfo{author}{\bibfnamefont{G.}~\bibnamefont{Weiglein}}, \bibnamefont{and}
  \bibinfo{author}{\bibfnamefont{K.~E.} \bibnamefont{Williams}}
  (\bibinfo{year}{2011}), \eprint{1102.1898}.

\bibitem[{\citenamefont{Abdo et~al.}(2010)\citenamefont{Abdo, Ackermann,
  Ajello, Atwood, Baldini et~al.}}]{Abdo:2010ex}
\bibinfo{author}{\bibfnamefont{A.}~\bibnamefont{Abdo}},
  \bibinfo{author}{\bibfnamefont{M.}~\bibnamefont{Ackermann}},
  \bibinfo{author}{\bibfnamefont{M.}~\bibnamefont{Ajello}},
  \bibinfo{author}{\bibfnamefont{W.}~\bibnamefont{Atwood}},
  \bibinfo{author}{\bibfnamefont{L.}~\bibnamefont{Baldini}},
  \bibnamefont{et~al.}, \bibinfo{journal}{Astrophys.J.}
  \textbf{\bibinfo{volume}{712}}, \bibinfo{pages}{147} (\bibinfo{year}{2010}),
  \eprint{1001.4531}.

\bibitem[{\citenamefont{Giedt et~al.}(2009)\citenamefont{Giedt, Thomas, and
  Young}}]{Giedt:2009mr}
\bibinfo{author}{\bibfnamefont{J.}~\bibnamefont{Giedt}},
  \bibinfo{author}{\bibfnamefont{A.~W.} \bibnamefont{Thomas}},
  \bibnamefont{and} \bibinfo{author}{\bibfnamefont{R.~D.} \bibnamefont{Young}},
  \bibinfo{journal}{Phys.Rev.Lett.} \textbf{\bibinfo{volume}{103}},
  \bibinfo{pages}{201802} (\bibinfo{year}{2009}), \eprint{0907.4177}.

\bibitem[{\citenamefont{Ahmed et~al.}(2009)}]{Ahmed:2009zw}
\bibinfo{author}{\bibfnamefont{Z.}~\bibnamefont{Ahmed}} \bibnamefont{et~al.}
  (\bibinfo{collaboration}{The CDMS-II}) (\bibinfo{year}{2009}),
  \eprint{0912.3592}.

\bibitem[{\citenamefont{Navarro et~al.}(1997)\citenamefont{Navarro, Frenk, and
  White}}]{Navarro:1996gj}
\bibinfo{author}{\bibfnamefont{J.~F.} \bibnamefont{Navarro}},
  \bibinfo{author}{\bibfnamefont{C.~S.} \bibnamefont{Frenk}}, \bibnamefont{and}
  \bibinfo{author}{\bibfnamefont{S.~D.} \bibnamefont{White}},
  \bibinfo{journal}{Astrophys.J.} \textbf{\bibinfo{volume}{490}},
  \bibinfo{pages}{493} (\bibinfo{year}{1997}), \eprint{astro-ph/9611107}.

\bibitem[{\citenamefont{Wilkinson}(2011)}]{bsmu_CMSLHCB}
\bibinfo{author}{\bibfnamefont{G.}~\bibnamefont{Wilkinson}}
  (\bibinfo{year}{2011}), \bibinfo{note}{talk presented at HEP-EPS 2011,
  Grenoble, July 2011.}

\bibitem[{\citenamefont{Paluton}(2011)}]{lhcb}
\bibinfo{author}{\bibfnamefont{M.}~\bibnamefont{Paluton}}
  (\bibinfo{year}{2011}), \bibinfo{note}{talk presented at Beauty 2011,
  Amsterdam, April 2011.}

\bibitem[{\citenamefont{Vivarelli}(2011)}]{ATLAS-susy}
\bibinfo{author}{\bibfnamefont{I.}~\bibnamefont{Vivarelli}}
  (\bibinfo{year}{2011}), \bibinfo{note}{talk presented at HEP-EPS 2011,
  Grenoble, July 2011.}

\bibitem[{\citenamefont{Conley et~al.}(2011)\citenamefont{Conley, Dreiner, and
  Wienemann}}]{Conley:2010jk}
\bibinfo{author}{\bibfnamefont{J.~A.} \bibnamefont{Conley}},
  \bibinfo{author}{\bibfnamefont{H.~K.} \bibnamefont{Dreiner}},
  \bibnamefont{and}
  \bibinfo{author}{\bibfnamefont{P.}~\bibnamefont{Wienemann}},
  \bibinfo{journal}{Phys. Rev.} \textbf{\bibinfo{volume}{D83}},
  \bibinfo{pages}{055018} (\bibinfo{year}{2011}), \eprint{1012.1035}.

\bibitem[{\citenamefont{Abazov et~al.}(2009)}]{Abazov:2009zi}
\bibinfo{author}{\bibfnamefont{V.}~\bibnamefont{Abazov}} \bibnamefont{et~al.}
  (\bibinfo{collaboration}{D0 Collaboration}), \bibinfo{journal}{Phys.Lett.}
  \textbf{\bibinfo{volume}{B680}}, \bibinfo{pages}{34} (\bibinfo{year}{2009}),
  \eprint{0901.0646}.

\bibitem[{\citenamefont{Forrest}(2009)}]{Forrest:2009gm}
\bibinfo{author}{\bibfnamefont{R.}~\bibnamefont{Forrest}}
  (\bibinfo{collaboration}{CDF Collaboration}) (\bibinfo{year}{2009}),
  \eprint{0910.1931}.

\end{thebibliography}
\end{document}